\documentclass[aps,twocolumn,amsmath,letterpaper]{revtex4}
\usepackage{amssymb}
\usepackage{graphicx}
\usepackage{array}
\usepackage{hhline}
\usepackage{longtable}
\usepackage{amsmath}

\renewcommand{\thetable}{\Roman{table}} \thetable

\begin{document}

\title{The Blume-Emery-Griffiths Spin Glass and Inverted Tricritical Points}
\author{V. Ongun \"Oz\c{c}elik$^1$ and A. Nihat Berker$^{2,3,4}$}
\affiliation{$^1$Department of Physics, Istanbul Technical
University, Maslak 34469, Istanbul, Turkey,}
  \affiliation{$^2$College of Sciences and Arts, Ko\c{c} University,
Sar\i yer 34450, Istanbul, Turkey,}
 \affiliation{$^3$Department of Physics, Massachusetts Institute of Technology, Cambridge,
Massachusetts 02139, U.S.A.,}
  \affiliation{$^4$Feza G\"ursey
Research Institute, T\"UB\.ITAK - Bosphorus University,
\c{C}engelk\"oy 34684, Istanbul, Turkey}

\begin{abstract}
The Blume-Emery-Griffiths spin glass is studied by
renormalization-group theory in $d=3$. The boundary between the
ferromagnetic and paramagnetic phases has first-order and two types
of second-order segments. This topology includes an inverted
tricritical point, first-order transitions replacing second-order
transitions as temperature is lowered. The phase diagrams show
disconnected spin-glass regions, spin-glass and paramagnetic
reentrances, and complete reentrance, where the spin-glass phase
replaces the ferromagnet as temperature is lowered for all chemical
potentials.

\end{abstract}

\maketitle
\def\s{\rule{0in}{0.28in}}

\setlength{\LTcapwidth}{\columnwidth}

\begin{figure}
\includegraphics[width=8cm]{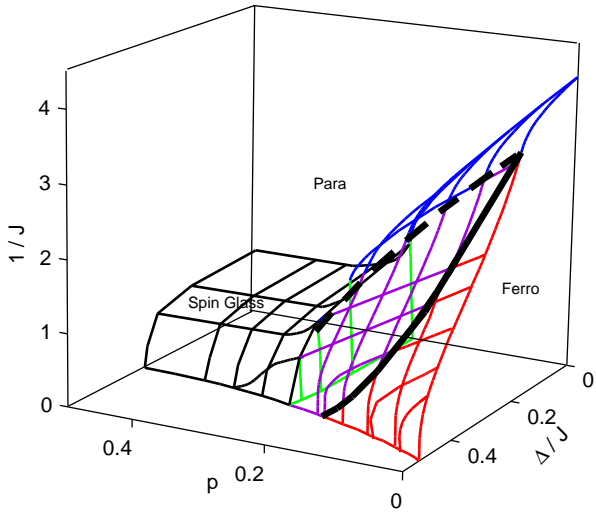}
\caption{(Color on-line) Our calculated global phase diagram for
$K=0$.  The ferromagnetic phase is bounded by a first-order surface
close to $p=0$, which recedes along the full line on the surface
from a new second-order transition induced by randomness and
controlled by a strong-coupling fixed distribution.  At the dashed
line, an ordinary second-order transition takes over.  The
transitions from the spin-glass phase, to the paramagnetic or
ferromagnetic phase, are second order. The system being symmetric
about $p=0.5$, the antiferromagnetic sector is not shown.}
\end{figure}

The Blume-Emery-Griffiths (BEG) model \cite{BEG,Berker} is the
simplest system for the study of the various meetings of first- and
second-order phase boundaries between ordered and disordered phases,
in a plethora of phase diagram topologies \cite{Hoston}.  In these
diagrams, the second-order phase transitions are dominated by
thermal fluctuations and occur at high temperatures.  The
first-order phase transitions evolve, to finite temperatures, from
zero-temperature ground-state energy crossings and occur at low
temperatures.  In a well-known phase diagram topology, a tricritical
point separates the high-temperature second-order boundary and the
low-temperature first-order boundary.  The BEG model has been used
to describe $^3$He-$^4$He mixtures \cite{BEG}, solid-liquid-gas
systems \cite{Lajzerowicz1}, multicomponent fluid and liquid-crystal
mixtures \cite{Lajzerowicz2}, microemulsions \cite{Schick},
semiconductor alloys \cite{Newman, Tanaka}, and electronic
conduction systems \cite{Kivelson}.

The inclusion of frozen disorder (quenched randomness) to these
systems should yield new phase diagrams, as is indeed seen in our
current work.  We find that a temperature sequence of transitions
that is reverse to the above can occur with the inclusion of
quenched randomness. Thus, an \emph{inverted tricritical point} is
obtained, separating a high-temperature first-order boundary and a
low-temperature second-order boundary. Since the BEG model is the
generic model for tricriticality, we believe that the quenched
randomness effect of inverted tricriticality should be quite
generally applicable.  With frustrated quenched randomness
\cite{NishimoriBook, Arenzon1, Arenzon2, Arenzon3}, the spin-glass
phase appears within the Blume-Emery-Griffiths global phase diagram
(Fig.1). Thus, a new spin-glass phase diagram topology is found, in
which disconnected spin-glass regions occur close to the
ferromagnetic and antiferromagnetic phases, but are separated by a
paramagnetic gap. No spin-glass phase diagram has to our knowledge
to-date yielded disconnected spin-glass regions. However, if the
global phase diagram of a physical realization of the BEG spin glass
is fully explored, these disconnected regions should be found.

We have studied, in spatial dimension $d=3$, the model with
Hamiltonian

\begin{equation}
-\beta \mathcal{H} = \sum_{<ij>}[
J_{ij}s_{i}s_{j}+Ks_{i}^2s_{j}^2-\Delta(s_{i}^2+s_{j}^2)],
\end{equation}

\noindent where $s_{i}=0,\pm1$ at each site $i$ of the lattice and
$<ij>$ indicates summation over nearest-neighbor pairs of sites. The
spin-glass type of quenched randomness is given by each local
$J_{ij}$ being ferromagnetic with the value $+J$ with probability
$1-p$ and antiferromagnetic with the value $-J$ with probability
$p$.  Under the scale change induced by renormalization-group
transformation, all renormalized interactions become quenched random
and the more general Hamiltonian

\begin{equation}
-\beta \mathcal{H} = \sum_{<ij>}[ J_{ij}s_{i}s_{j}
+K_{ij}s_{i}^2s_{j}^2
-\Delta_{ij}(s_{i}^2+s_{j}^2)-\Delta_{ij}^\dagger(s_{i}^2-s_{j}^2)],
\end{equation}

\noindent has to be considered.  The renormalization-group flows are
in terms of the joint quenched probability distribution $P(J_{ij},
K_{ij}, \Delta_{ij}, \Delta_{ij}^\dagger)$, which is renormalized
through the convolution \cite{Falicov}

\begin{equation}
P'(\textbf{K}'_{i'j'})=\int[\prod_{ij}^{i'j'}d\textbf{K}_{ij}P(\textbf{K}_{ij})]\delta(\textbf{K}'_{i'j'}-
\textbf{R}(\{\textbf{K}_{ij}\})),
\end{equation}

\noindent where primes refer to the renormalized system,
$\textbf{K}_{ij} \equiv (J_{ij}, K_{ij}, \Delta_{ij},
\Delta_{ij}^\dagger)$, and $\textbf{R}({\textbf{K}_{ij}})$ is the
local recursion relation through which 108 unrenormalized local
interactions in $\{\textbf{K}_{ij}\}$ determine 4 renormalized local
interactions in $\textbf{K}'_{i'j'}$.  The local recursion relation
$\textbf{R}({\textbf{K}_{ij}})$ is effected by a mixed
Migdal-Kadanoff procedure \cite{Erbas} with $d=3$ and length
rescaling factor $b=3$ necessary for the equal \textit{a priori}
treatment of ferromagnetism and antiferromagnetism.  Thus, our
treatment is approximate for the cubic lattice and exact for the
hierarchical lattice
\cite{BerkerOstlund,Kaufman,Kaufman2,Erbas,Hinczewski2,Hinczewski3,ZRZ,
ben-Avraham,Khajeh,Garel} shown in Fig.2.  This hierarchical lattice
is known to give very accurate results for the critical temperatures
of the d =3 isotropic and anisotropic Ising models on the cubic
lattice \cite{Erbas}. Furthermore, in general, exact calculations
for hierarchical lattices have been seen to constitute very good
approximations for cubic
lattices.\cite{Falicov,Garel,Migliorini,Nobre,Hui}

The probability distribution $P(J_{ij}, K_{ij}, \Delta_{ij},
\Delta^{\dagger}_{ij})$ is represented by histograms lodged on a
four-dimensional interaction space $(J_{ij}, K_{ij}, \Delta_{ij},
\Delta^{\dagger}_{ij})$.  Eq.(3) is effected by 8 pairwise
convolutions, which are either bond-moving or decimation in the
appropriate sequence, between intermediate distributions. The number
of histograms rapidly grows from the starting two described after
Eq.(1). Thus, for calculational purposes, before each pairwise
convolution, the histograms are combined by using a binning
procedure, so that our results are obtained by the
renormalization-group flows of 22,500 histograms.

\begin{figure}
\includegraphics[width=6cm]{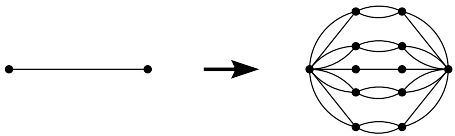}\\
\caption{The $d=3$ hierarchical lattice for which our calculation is
exact is constructed by the repeated imbedding of the graph as shown
in this figure. This hierarchical lattice gives very accurate
results for the critical temperatures of the d =3 isotropic and
anisotropic Ising models on the cubic lattice \cite{Erbas}.}
\end{figure}

Tricritical phase diagram cross-sections of the purely ferromagnetic
system for different $K/J$ values are shown in Fig.3.  These are
standard tricritical phase diagrams, in the absence of quenched
randomness, with the tricritical point separating the second-order
transitions at high temperature and the first-order transitions at
low temperature.  The humped boundary, occurring in mean-field
theory but not in the $d=2$ system \cite{Berker}, is thus found to
occur in the $d=3$ system.

\begin{figure}
\includegraphics[width=8cm]{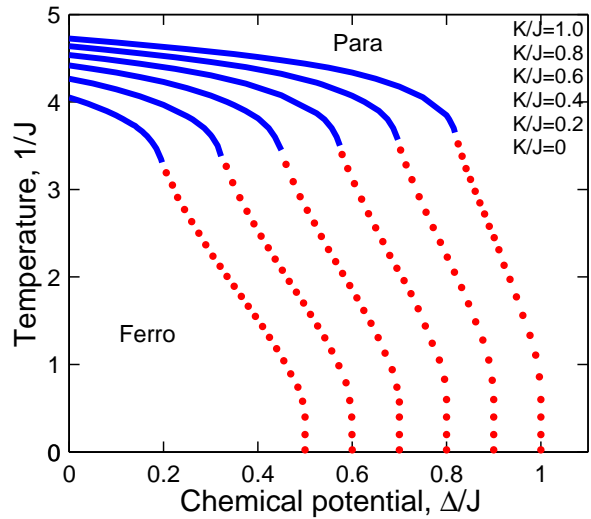}
\caption{(Color on-line) Tricritical phase diagram cross-sections of
the purely ferromagnetic system for the $K/J$ values indicated on
the figure, shown consecutively from the innermost curve for $K/J =
0$. First- and second-order transitions are shown by dotted and full
lines, meeting at a tricritical point. In these systems, with no
quenched randomness, the standard tricritical topology occurs, with
the second-order boundary at high temperature and the first-order
boundary at low temperature.}
\end{figure}

Our calculated global phase diagram for the BEG spin-glass system is
in Fig.1 for $K=0$.  The ferromagnetic phase is bounded by a
first-order surface close to $p=0$, which recedes along the full
line on the surface from a new second-order transition induced by
randomness and controlled by a strong-coupling fixed distribution.
At the dashed line, an ordinary second-order transition takes over.
The full line is thus a line of random-bond tricritical points.  The
dashed line is a line of special critical points around which
universality is violated, since the second-order phase transitions
on each side of this line have different critical
exponents.\cite{Falicov}  These two lines meet at the non-random
$(p=0)$ tricitical point.  The transitions from the spin-glass
phase, to the paramagnetic or ferromagnetic phase, are second order.

\begin{figure}
\includegraphics[width=8cm]{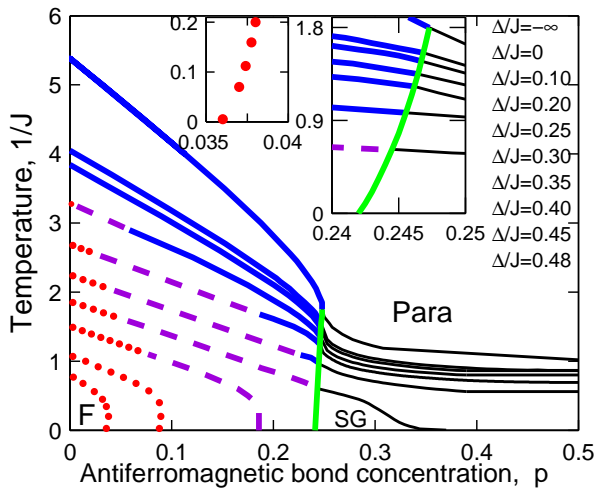}
\caption{(Color on-line) Blume-Emery-Griffiths spin-glass phase
diagrams: Constant $\Delta/J$ cross-sections of the global phase
diagram in Fig.1. The outermost cross-section has
$\Delta/J=-\infty$, meaning no $s_i=0$ states. The successive
cross-sections, going inwards from the outermost cross-section, are
for the successively higher values of $\Delta/J$ indicated on the
figure.  Thus, annealed vacancies $s_i=0$ are introduced in these
cross-sections with successively higher values of $\Delta/J$, making
all ordered phases recede.  The dotted and full lines are
respectively first- and second-order phase boundaries. The dashed
lines are strong-coupling second-order phase boundaries induced by
quenched randomness.  The inverted tricritical topology is seen
between the dotted and dashed lines, with the first-order
transitions occurring at high temperature and the second-order
transitions occurring at low temperature, on each side of the
tricritical point.  A new spin-glass phase diagram topology is
obtained for $\Delta/J=0.35$, in which the spin-glass phase occurs
close to the ferromagnetic (and, symmetrically, antiferromagnetic,
not shown here) phase, but yields to the paramagnetic phase as $p$
is increased towards 0.5. The spin-glass phase disappears at
$\Delta/J=0.37$.  The insets, with expanded scales, clearly show the
ferromagnetic to paramagnetic phase reentrance and the ferromagnetic
to spin-glass phase reentrance.}
\end{figure}

\begin{figure}
\includegraphics[width=8cm]{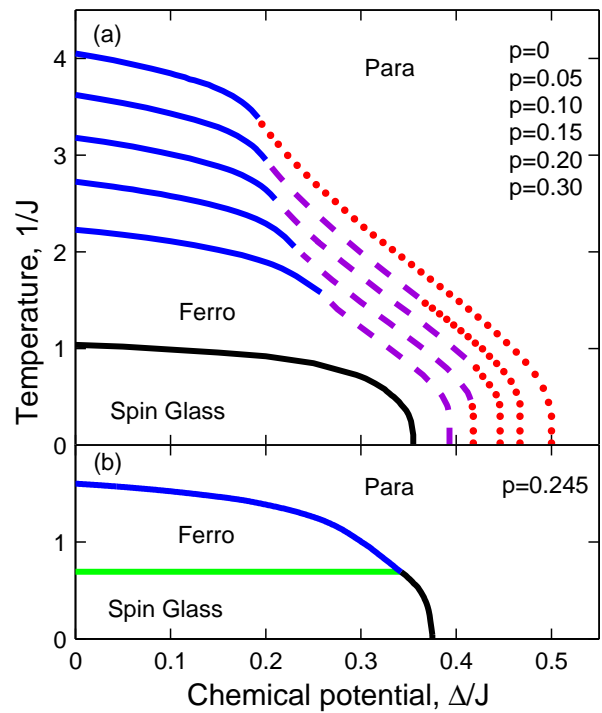}
\caption{(Color on-line) Spin-glass Blume-Emery-Griffiths phase
diagrams: Constant $p~$ cross-sections of the global phase diagram
in Fig.1. The dotted and full lines are respectively first- and
second-order phase boundaries.  The dashed lines are strong-coupling
second-order phase boundaries induced by quenched randomness.  The
outermost curve in (a) corresponds to the pure Blume-Emery-Griffiths
model with no quenched randomness $(p=0)$.  The successive
cross-sections in (a), going inwards from the outermost
cross-section, are for the successively higher values of $p~$
indicated on the figure.  As spin-glass quenched randomness is
introduced with increasing values of $p$, ordered phases and
first-order phase transitions recede.}
\end{figure}

Cross-sections of this global phase diagram for constant chemical
potential $\Delta/J$ of the non-magnetic state are in Fig.4. The
outermost cross-section has $\Delta/J=-\infty$, meaning no $s_i=0$
states, and therefore is the phase diagram of the spin-1/2 Ising
spin glass, showing as temperature is lowered the
paramagnet-ferromagnet-spinglass reentrance \cite{Migliorini,
Nobre}. The annealed vacancies, namely the nonmagnetic states
$s_i=0$, are introduced in cross-sections with successively higher
values of $\Delta/J$. For $\Delta/J$ greater than the non-random
tricritical value of $\Delta/J = 0.192$, first-order transitions
between the ferromagnetic and paramagnetic phases are introduced
from the low randomness side, but are converted to the
strong-coupling second-order transition at a threshold value of
randomness $p$. This constitutes an \emph{inverted tricritical
point}, since the phase boundary is converted from first order to
second order as temperature is lowered, contrary to the ordinary
tricritical points (as seen for example in Fig.3).  The above
results are consistent with the general prediction that, in $d = 3$,
quenched randomness converts first-order boundaries into second
order, at a threshold amount of randomness.\cite{Hui}  (In $d = 2$,
this conversion is predicted to happen with infinitesimal quenched
randomness.\cite{Hui, Aizenman}).  The randomness threshold in $d =
3$ is of course higher for stronger first-order transitions.  In the
current system, increased frustrating quenched randomness has two
parallel effects, namely driving the phase transition to lower
temperature and reaching the threshold for the conversion to second
order, which explains the calculated results of inverted
tricriticality. (With non-frustrating quenched randomness, on the
other hand, the transition can actually be driven to higher
temperature, while the conversion to second order still happens
\cite{Malakis}.)

\begin{figure*}
\includegraphics*[scale=1.8]{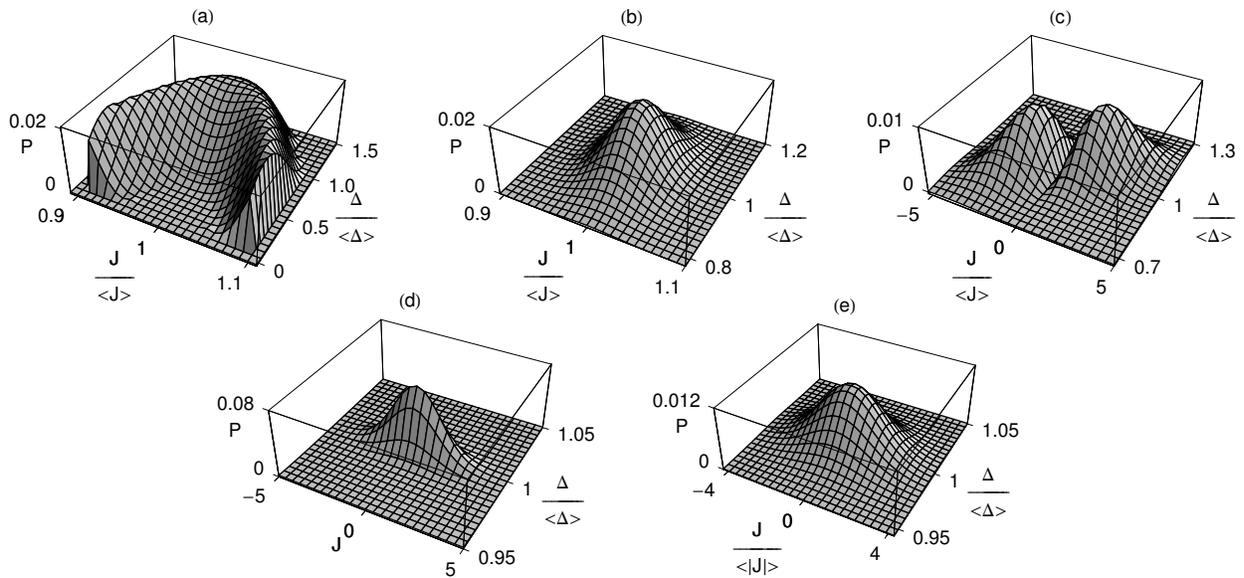}
\caption{Projections of the fixed distributions $P^*(J_{ij}, K_{ij},
\Delta_{ij}, \Delta^{\dagger}_{ij})$ for: (a) the disorder-induced
second-order transitions between the ferromagnetic and paramagnetic
phases, (b) the first-order transitions between the ferromagnetic
and paramagnetic phases, (c) the second-order transitions between
the ferromagnetic and spin-glass phases, (d) the second-order phase
transitions between the spin-glass and paramagnetic phases, and (e)
the sink fixed distribution for the spin-glass phase. Note that
(a),(b),(c),(e) are runaways, in the sense that the couplings
renormalize to infinity while the distribution retains its shape
shown here.  In the second-order phase transitions between the
spin-glass and paramagnetic phases (d), $\Delta$ is a runaway (to
minus infinity), while the other interactions remain finite. The
fixed distributions in this figure are singly unstable, except for
the sink (e), which is totally stable.}
\end{figure*}

As the annealed vacancies $s_i=0$ are increased, at $\Delta/J \geq
0.34$, of the second-order transitions between the ferromagnetic and
paramagnetic phases, only the strong-coupling transition remains. At
$\Delta/J \geq 0.42$, the strong-coupling second-order transition
also disappears, leaving only first-order transitions between the
ferromagnetic and paramagnetic phases.  Also as the annealed
vacancies are increased, all ordered phases recede.  In this
process, first the spin-glass phase disappears, at $\Delta/J =
0.37$, which is understandable, since it is tenuously ordered due to
frustration. The new, \emph{disconnected spin-glass} phase diagram
topology is obtained in this neighborhood, e.g., for $\Delta/J=0.35$
as shown in Fig.4, in which the spin-glass phase occurs close to the
ferromagnetic (and, symmetrically, antiferromagnetic, not shown in
the figures) phase, but yields to the paramagnetic phase as $p$ is
increased towards 0.5 .  The disconnected spin-glass phase diagrams
occur in a very narrow portion of the global phase diagram.

The paramagnetic-ferromagnetic-spinglass reentrances, as temperature
is lowered, of the Blume-Emery-Griffiths spin-glass cross-sections
fall on the same reentrant second-order boundary, as seen in Fig.4.
As seen for $\Delta/J = 0.45$ and 0.48 in this figure, before
disappearing at $\Delta/J = 0.5$, the ferromagnetic phase exhibits
paramagnetic-ferromagnetic-paramagnetic reentrance as temperature is
lowered.

Constant $p$ cross-sections of the global phase diagram in Fig.1 are
shown in Fig.5.  The outermost curve corresponds to the pure
Blume-Emery-Griffiths model with no quenched randomness $(p=0)$. As
spin-glass quenched randomness is introduced with increasing values
of $p$, we see that the first-order boundary recedes to the
strong-coupling second-order boundary, while the ordinary
second-order boundary also expands. At $p = 0.18$, the first-order
transition completely disappears.  At $p = 0.241$, the spin-glass
phase appears below the ferromagnetic phase, reflecting complete
reentrance. At $p = 0.249$, the spin-glass phase completely replaces
the ferromagnetic phase as the ordered phase, which is enveloped by
second-order transitions only.  Thus, for $0.241 < p < 0.759$, the
second-order boundary between the spin-glass and paramagnetic phases
reaches zero temperature.

In the results above, the phase diagrams are determined by the
basins of attraction of the renormalization-group sinks, namely the
completely stable fixed points and fixed distributions:  Each basin
is a thermodynamic phase.  The nature of the phase transitions is
determined by analysis of the unstable fixed points and fixed
distributions to which the phase diagram points of these transitions
flow.  Fig.6 shows the unstable fixed distributions of (a) the
quenched randomness-induced second-order transitions between the
ferromagnetic and paramagnetic phases, (b) the first-order
transitions between the ferromagnetic and paramagnetic phases, (c)
the second-order transitions between the ferromagnetic and
spin-glass phases, and (d) the second-order transitions between the
spin-glass and paramagnetic phases. The (totally stable) sink fixed
distribution of the spin-glass phase is also shown, in (e). The
eigenvalue exponent of the unstable fixed distribution controlling
(b) the first-order transitions between the ferromagnetic and
paramagnetic phases is $y = 3 = d$, as is required for first-order
transitions \cite{Nienhuis}. The eigenvalue exponents of the other
unstable fixed distributions, (a),(c),(d), are $y < d$ as is
required for second-order transitions.

We thank C. G\"uven M. Hinczewski, C.N. Kaplan, and O.S. Sar\i yer
for comments.  This research was supported by the Scientific and
Technological Research Council (T\"UB\.ITAK) and by the Academy of
Sciences of Turkey.

\end{document}